# ГЕОМЕТРИЧЕСКОЕ МОДЕЛИРОВАНИЕ ДЕНДРИТНЫХ СТРУКТУР

**Александр С. Прохода, к.ф.-м.н.**

Днепр, Украина

Приведен рецепт качественного, кинетического моделирования геометрическими методами трехмерных дендритных кристаллов. Установлены характерные особенности возмущений появляющихся на поверхности шарообразного тела, приводящие к различным сценариям формирования одной и той же равновесной формы кристалла. Выполнены компьютерные реализации таких дендритных кристаллов растущих как в условиях без конвективного потока, так и в набегающем потоке жидкости. Особое внимание уделено топологическим характеристикам моделей с некристаллографическими точечными группами симметрии. Исследованы поверхности разного рода, в том числе и 270.

## 1. ОСНОВЫ ГЕОМЕТРИЧЕСКОГО МОДЕЛИРОВАНИЯ ТРЕХМЕРНЫХ ДЕНДРИТНЫХ КРИСТАЛЛОВ

Однородная среда характеризуется инвариантностью физического свойства по отношению к трансляции. В изотропной среде физическое свойство инвариантно относительно вращения вокруг неподвижной точки. Анизотропная среда характеризуется зависимостью измеряемого свойства от направления измерения. Удобно изображать зависимость свойства от направления, по которому это свойство определяется, моделью, в которой величина свойства равна расстоянию от центра модели до ее поверхности.

---

E-mail: a-prokhoda@mail.ru

Поверхностная энергия γ кристаллов анизотропна. Зависимость γ от ориентации поверхности изображают при помощи полярных диаграмм. Зная полярную диаграмму, можно воссоздать равновесную форму кристалла. Обратно, из экспериментальных данных о равновесной форме кристалла можно построить соответствующую полярную диаграмму. Согласно с Херингом (см. например [1]), равновесную форму кристалла можно восстановить из полярной диаграммы поверхностной энергии как внутреннюю, выпуклую огибающую семейство плоскостей, касательных к полярной диаграмме в каждой её точке. Если, на полярной диаграмме имеются сингулярные точки (минимумы), то на равновесной форме кристалла обязательно присутствуют грани, которые отвечают этим сингулярным точкам.

Отметим, что эксперименты по определению равновесной формы непросты т.к. кристалл необходимо длительное время выдерживать в равновесии со средой. Габитус кристалла определяется его совокупностью граней. Равновесная форма может слагаться из граней и закругленных участков между ними. При увеличении температуры часто происходит разупорядочение поверхностной структуры, поверхность кристалла становится шероховатой в атомном масштабе. При этом изменяется и полярная диаграмма, на её минимумах исчезают сингулярные точки. Соответственно, на равновесной форме появляются закругленные участки. При дальнейшем повышении температуры площадь граней уменьшается и со временем исчезает полностью. Равновесная форма становится шарообразной. А при отклонении от равновесия кристаллы растут в виде закругленных дендритов.

Согласно принципу Кюри-Вульфа, свободная поверхностная энергия для кристалла, который находится в равновесии с окружающей средой, должна быть минимальной $\sum_i \gamma_i \omega_i = 0$, где $\omega_i$ – площадь $i$-ой грани. Таким образом, чем меньше поверхностная энергия *i*-ой грани, тем ближе эта грань до центра кристалла и тем больше её площадь по сравнению с другими гранями. Поверхностная энергия минимальна для граней, которые отвечают наиболее плотно упакованным атомами кристаллическим плоскостям. Эти грани являются основными в габитусе кристалла. При увеличении температуры (например, при движении вверх вдоль линии ликвидуса при изменении состава бинарных расплавов вблизи основного компонента), появляются дополнительные грани и постепенно увеличивается их относительная площадь. И со временем могут появиться закругленные участки. Из чистых расплавов металлов преимущественно растут шарообразные кристаллы, а огранка появляется при понижении температуры и конкретном составе расплава, связанных между собой условием равновесия – линия ликвидуса на диаграмме состояния. Кристаллы чистых металлов имеют наиболее часто шарообразную форму, что свидетельствует о значительной шероховатости в атомном масштабе межфазной поверхности раздела кристалл-расплав.

Способность кристалла самоограняться заключается в том, что при подходящих условиях растущий кристалл приобретает форму многогранника с плоскими гранями. Форма кристалла в сильной степени зависит от условий роста. С макроскопической, классической точки зрения кристаллом может быть названо однородное анизотропное тело, способное самоограняться [2]. Однако для характеристики подкласса класса кристаллов, который составляют квазикристаллы в данном определении свойство однородности необходимо заменить на свойство неоднородности.

Говорят, что рост кристалла анизотропен, если под действием данной движущей силы (например, переохлаждения [3]) разные грани растут с неодинаковыми скоростями. Если различие скоростей роста граней по сравнению с температурными градиентами вблизи фазовой границы невелики, то граница с хорошим приближением изотермична и перемещается по нормали к изотермам в системе. В случае же большей анизотропии скорости роста граница далеко не изотермична, т.е. на разных ее участках, растущих с одинаковой скоростью, температура может существенно различаться. Следует отметить, что нормальному механизму соответствует изотропный (или почти изотропный) рост, а послойному – анизотропный. Обратное, вообще говоря, неверно: если обнаружена анизотропия, то, прежде чем связывать её с механизмом роста, следует учесть возможность существования в системе температурных градиентов. В работе [3] также рассмотрен критерий Джексона, согласно которому, зная энтропию плавления, кристаллическую структуру и тип химической связи, можно предсказывать морфологию роста.

Далее перейдем к рассмотрению устойчивости форм роста. Итак, кристаллы с выпуклыми округлыми или гранными формами разрастаются, оставаясь подобными себе, лишь в определенных условиях. Например, в неперемешивающейся маточной среде подобное разрастание возможно, пока размер кристалла и отклонения от равновесия не превышают определенных значений [4, 5]. В противном случае, кристаллы приобретают скелетную или дендритную форму. Дендрит развивается из каждого ствола скелета в результате появления ветвей второго, третьего и т.д. порядков. Скелеты и дендриты возникают вследствие неустойчивой начальной выпуклой формы кристалла по отношению к случайно возникающим возмущениям. Физической причиной, которая обуславливает появление и развитие возмущений той или иной симметрии, является анизотропия скорости роста кристалла. Симметрия угловой зависимости кинетического коэффициента слегка отличается от сферической даже для шероховатой в атомном масштабе поверхности. На начальной стадии потери стабильности формы роста выступы, связанные с анизотропией скорости роста, начинают опережать иные области межфазной границы. Дальнейшее их перемещение в пересыщенный раствор приводит к возникновению дендритов. Каждый ствол развитого скелета растет уже независимо от первичных выступов и имеет форму, близкую к параболоиду, причем скорость его роста определяется кривизной вершины параболоида. На определенном расстоянии от вершины радиус кривизны поверхности параболоида становится больше критического $R_c$, и эта поверхность также теряет устойчивость – образуются боковые ветви дендрита.

Процесс роста кристалла включает в себя не только поверхностные процессы – необходимой стадией является также доставка кристаллизующихся компонентов к фазовой границе и отвод теплоты кристаллизации. При росте из конденсированных фаз – растворов и расплавов, а также из достаточно плотных газов перенос чаще всего осуществляется с участием принудительного перемешивания или естественной конвекции. Однако у твердых поверхностей, в том числе кристаллических, существует неподвижный приграничный слой, в котором перенос можно рассматривать как происходящий путем обычной диффузии или теплопроводности. Теория процессов переноса при кристаллизации рассмотрена в работах Любова (см. например [6]). Обычно совершенство кристалла уменьшается с увеличением пересыщения на растущей поверхности. Поэтому качество кристалла может служить грубой мерой пересыщения. Выделение теплоты кристаллизации повышает температуру поверхности и тем самым снижает скорость роста. Однако, поскольку коэффициент диффузии обычно примерно на три порядка меньше коэффициента температуропроводности, указанный эффект важен для систем, в которых составы кристалла и жидкой фазы близки друг к другу по обоим компонентам. Более интенсивный теплоотвод от вершин кристалла, чем от его граней, определяет разницу температур вдоль грани. Неэквивалентные условия доставки вещества и теплоотвода около вершин кристалла, его ребер и в центрах граней являются общими для любых растущих многогранников. При росте из растворов окрестности вершин питаются из большего телесного угла, и поэтому пересыщение вблизи от них на поверхности кристалла выше, чем над центральными участками граней. Аналогично, при росте изолированного кристалла в расплаве переохлаждение около его вершин наибольшее по сравнению с другими участками фронта роста. Такое распределение – одна из главных причин генерации слоев в окрестностях вершин и ведет к образованию скелетных и дендритных форм роста.

Форма, которую кристалл принимает в процессе образования – форма роста – очень чувствительна к условиям кристаллизации и так же, как и морфология поверхности, отражает механизм роста. Поэтому формы роста позволяют судить об условиях образования и корректировать параметры кристаллизации, если речь идет об искусственном выращивании кристаллов.

Полиэдральная форма возникает, очевидно, когда каждая грань нарастает, оставаясь параллельно самой себе, т.е. когда скорость роста зависит только от ориентации поверхности и одинакова (в среднем) для всех ее точек: $v = v(\mathbf{n})$. Чернов в работе [4] показал, что стационарная форма роста в описанных выше условиях определяется по правилу Вульфа с помощью функции $v(\mathbf{n})$ точно так же, как равновесная форма с помощью удельной поверхностной энергии $\alpha(\mathbf{n})$. Превращение атомно-гладкой поверхности в шероховатую означает округление острого минимума зависимости $v(\mathbf{n})$. Размер той или иной грани или протяженность округлого участка с тем или иным набором ориентаций на форме роста определяется в соответствии с $(\mathbf{n}, \mathbf{r}) = v(\mathbf{n}) \cdot t$ т.е. аналогом правила Кюри-Вульфа, где $\alpha(\mathbf{n}) \to v(\mathbf{n})$, $2\Omega/\Delta\mu \to t$, и $t$ – время от момента зарождения кристалла. Из геометрической процедуры построения огибающей Вульфа следует, что чем выше скорость роста грани (или округлого участка), тем меньше должен быть её размер на стационарной форме. Структура поверхности и механизм роста существенно зависят от пересыщения, температуры и состава окружающей среды, и потому габитус кристалла зависит не только от его структуры, но и от упомянутых факторов. Растущий кристалл создает вокруг себя некоторое распределение температуры, которое в свою очередь определяет переохлаждение на фронте роста в каждой его точке и, следовательно, скорость роста и форму кристалла. В этих условиях стационарная, не меняющаяся во времени форма роста кристалла возможна лишь, если фронт роста представляет собой плоскость, эллипсоид или параболоид. Этот теоретический результат был доказан Иванцовым в роботе [7] который справедлив при бесконечно быстрой поверхностной кинетике и не учитывает смещения температуры равновесия из-за кривизны фронта роста.

Одной из замечательных особенностей процесса кристаллизации из расплава, является различие в поведении металлов и неметаллов. Различие определяется строением фазовой границы (кристалл-расплав) в атомном масштабе: у металлов эта граница шероховатая, у большинства неметаллов – гладкая. Критерием различия служит энтропия фазового превращения, численное значение которой для шероховатых поверхностей меньше по сравнению с гладкими поверхностями. Отметим, что большинство органических и неорганических материалов, у которых энтропия плавления принимает значения из интервала $(2 < H_{пл}/kT_e < 4)$ в чистом виде кристаллизуются с огранением, но при достаточном количестве примесей ограняются по типу псевдодендритов. Вершины псевдодендритов закруглены. В общем случае кристалл ограняется, если при данном градиенте температур устанавливается определенное соотношение между разностью температур на разных участках некоторой грани (все они растут с одинаковой скоростью) и расстоянием между этими участками. Таким образом, на вершине, где градиенты велики, кристалл оказывается закругленным (не ограненным). Боковые же участки кристалла растут медленнее, но градиенты там меньше, поэтому они покрываются фасетками. Таким образом, данного рода кристаллы можно заставить расти как в виде псевдодендритов так и сферолитов.

Помимо чисто гранных и чисто округлых форм возможна комбинация тех и других на одном кристалле. В зависимости от соотношения осевого и радиального тепловых потоков изотермы в кристалле могут быть выпуклыми или вогнутыми, следовательно, имеем выпуклый или вогнутый фронт роста.

Таким образом, переохлаждение на фронте кристаллизации и анизотропия роста кристаллов в общем случае тем больше, чем больше энтропия плавления. На скорость роста существенно влияет также диффузионная подвижность атомов. Однако именно тип связи определяет анизотропию роста, поскольку от него зависят скорости присоединения и отрыва атомов (молекул) кристалла.

Опыт показывает, что совершенные гранные формы кристаллов возникают, когда пересыщение или переохлаждение на фронте роста достаточно мало. С увеличением отклонения от равновесия кристалл меняет свой облик, превращаясь в скелет или дендрит. Чернов в работе [4] рассмотрел схему преобразования совершенного кристалла в скелет. На первой стадии превращения кристалл, продолжает оставаться многогранником, но при этом появляются прогибы граней с максимальными углами при их центрах. Потом в центрах граней возникают впадины, где собирается примесь, и признаки сплошности грани теряются. Дальнейшее повышение пересыщения ведет к образованию скелетных форм. Наконец, при ещё бóльших пересыщениях возникают дендриты. Отметим, что скелетный рост обусловлен непостоянством пересыщения вдоль граней. Чем больше размер кристалла при сравнении с отношением $D/\beta$, где $D$ – коэффициент диффузии, $\beta$ – кинетический коэффициент, тем больше должны искривляться его грани, чтобы компенсировать неоднородность пересыщения. Появление прогиба в центре граней при больших пересыщениях ведет к дальнейшему ухудшению питания этих участков и, следовательно, к еще большему отставанию их от вершин. Искривления приводят к появлению на поверхностях участков с большим значением кинетического коэффициента, однако при этом пропадает его анизотропия. В результате этих изменений искривления увеличиваются все больше и так далее. Таким образом, достижение некоторых критических значений локальных прогибов ведет к лавинообразной потере устойчивости, а точнее – к невозможности роста многогранника подобным к самому себе.

В работе [8] приведен обзор теорий устойчивости форм роста кристаллов и соответствующие численные методы расчета эволюции таких форм. В современных работах, например [9, 10] для решения задач формообразования кристаллов используют решеточный метод, основанный на кинетическом уравнении Больцмана. Использование очень мелких решеток (сеток – в двумерном случае) позволяет точно определить положение межфазной поверхности в модели фазового поля, рассчитать локальный наклон поверхности и ее кривизну, что необходимо для учета анизотропии поверхностной энергии и кинетического коэффициента. Такое моделирование, возможно, произвести например, на современных видеоадаптерах применяя технологию CUDA. Однако, главной проблемой методик, основанных на кинетическом уравнении Больцмана, являются трудности подбора коэффициентов, которые характеризуют акты рассеяния выбранных элементов массы, таким образом, чтобы результаты моделирования отвечали табличным значениям коэффициентов диффузии, теплопроводности и т.д. Более того, для реализации данной методики моделирования используются решетки в классическом понимании (в смысле Федорова). Данные решетки пригодны для моделирования кристаллов обладающих поворотными симметриями только таких порядков: 1, 2, 3, 4 и 6. Так как расчет происходит по узлам решетки, то очевидно, что данные узлы «навязывают» соответствующие симметрии. Следовательно, такие решетки не пригодны для моделирования квазикристаллов скажем с икосаэдральной группой симметрий. Для моделирования квазикристаллов в модели фазового поля необходимы квазирешетки!

Далее вспомним о некоторых топологических характеристиках объектов.

Со времен Мёбиуса было известно, что каждую замкнутую поверхность можно построить, добавив к сфере $p$ ручек и $q$ скрещенных колпаков («пленок Мёбиуса»). Напомним, что знакопеременная сумма чисел граней, ребер и вершин полиэдральной поверхности называется её эйлеровой характеристикой $\chi$. Для замкнутых поверхностей имеет место следующее соотношение $\chi = F - E + V = 2 - 2p - q$, где $F, E, V$ – соответственно число граней, число рёбер и число вершин. Род $p = (2 - \chi)/2$ однозначно характеризует ориентируемые (двусторонние) поверхности (к которым относятся, например, все римановы поверхности), с точностью до топологической эквивалентности. В этом случае $q = 0$, поскольку наличие ленты Мёбиуса на поверхности сделало бы её односторонней. Связное многообразие $N$ ориентировано тогда и только тогда, когда параллельный перенос вдоль любого замкнутого пути (начинающегося и кончающегося в одной точке) сохраняет ориентацию. Двухсторонние поверхности представлены: сферой $S^2$ (род которой ноль); тором $T^2$ (рода один); двудырым тором, кренделем (рода два); трёхдырым тором (рода три) и т.д. Все они могут быть вложены в трехмерное пространство. Отметим, что в евклидовой и эллиптической геометрии существует только по одной замкнутой двусторонней пространственной форме (именно кольцо и сфера), в гиперболической геометрии существует бесчисленное множество замкнутых двусторонних, двумерных пространственных форм. Согласно Клейну [11] всякая поверхность рода $p \geq 2$ может быть рассматриваема как гиперболическая пространственная форма, т.е. что её можно снабдить всюду регулярным гиперболическим мероопределением совершенно аналогично тому, как в случае поверхности рода $p = 1$ мы имеем евклидову пространственную форму.

С замкнутой поверхностью связаны две важные группы: фундаментальная группа и группа классов преобразований. Фундаментальную группу, именуемую также, первой гомотопической группой поверхности (рассматриваемой как топологическое пространство) определил и исследовал Пуанкаре в работе [12].

В случае фундаментальной группы выберем на поверхности отмеченную точку и рассмотрим множество всех непрерывных путей, которые начинаются и кончаются в этой точке. Возможность продеформировать одну такую петлю в другую позволяет задать отношение эквивалентности между петлями, согласованное с бинарной операцией на петлях (произведение двух петель получается, если соединить конец первой с началом второй). Для того чтобы пройти по поверхности вдоль петли-произведения, нужно, обойдя первую петлю, сразу же отправится в путешествие по второй. Единичный элемент получающейся группы представляется теми петлями, которые можно, стянуть в точку (петлями гомотопными нулю). Если, например, в конце путешествия повернуть назад и пройти свой путь в обратном направлении, то такое произведение данной петли и ей обратной можно стянуть в точку, не сходя с исходного пути. В общем случае гомотопная нулю петля может заметать какой-то участок поверхности при стягивании ее в отмеченную точку.

Заметим, что гомотопический класс произведения двух направленных путей не изменится, если заменить сомножители гомотопными. Поэтому можно говорить о произведении гомотопических классов направленных путей. Во многих случаях открытое многообразие $M$ размерности $n$ можно продеформировать по себе к подмножеству меньшей размерности, для которого вычисление фундаментальной группы и других инвариантов значительно проще. Фундаментальная группа неориентируемого многообразия нетривиальна и имеет ненулевой гомоморфизм $\sigma$ в группу из двух элементов.

На сфере нет ручек, к которым можно было бы привязать лассо, а потому и сама фундаментальная группа тривиальна. На любой другой поверхности имеется, по крайней мере, одна ручка или пленка Мёбиуса, за которую «зацеплен» нетривиальный элемент фундаментальной группы. Пуанкаре сформулировал (гипотезу) теорему, что фундаментальная группа вместе со своими многомерными аналогами всегда дает возможность отличить сферы от более сложных многообразий. Перельману [13] доказал данную теорему для случая трехмерной сферы.

Фундаментальную группу поверхности можно считать алгебраической записью трудностей, с которыми сталкиваемся при попытке стянуть петлю к отмеченной точке. По аналогии группа классов преобразований поверхности – это алгебраическая запись препятствий к деформации преобразования в тождественное отображение. При тождественном отображении каждая точка поверхности переходит в саму себя. В частности, группа классов преобразований изоморфна группе внешних автоморфизмов фундаментальной группы поверхности. Отметим, что ряд важных топологических идей к реализации такого моделирования мы почерпнули из книги Франсиса [14].

Далее перейдем непосредственно к результатам.

Отметим, что такой геометрический подход является не статическим, а кинетическим, т.е. из исходной, начальной модельной формы тела путем дискретных деформаций следует следующая, а потом последующая и т.д. Причем интервал дискретность можно устремить к нулю, что позволяет моделировать непрерывные деформации (т.е. осуществлять гомеоморфизмы).

Обратим внимание, что мы не преследуется решение задачи установления количественных, подгоночных параметров процесса кристаллизации в предлагаемой модели и сравнения их с известными экспериментальными данными, а производим лишь компьютерные реализации некоторых из возможных форм дендритных квазикристаллов.

Далее по тексту анализируются поверхности дендритных квазикристаллов при этом подразумевается, что данные дендриты являются телами.

## 2. АНАЛИЗ РЕЗУЛЬТАТОВ МОДЕЛИРОВАНИЯ

Наделения неэвклидовой (гиперболической) геометрией большинства трехмерных многообразий, изложены в обзорных статьях Тёрстона [15] и Милнора [16]. В схеме Тёрстона именно топология многообразия ограничивает и часто определяет его возможные геометрии. Главный способ получить геометрическое многообразие заключается в том, чтобы взять кусочно-гладкую область геометрического пространства, имеющий форму многогранника, и попарно отождествить между собой его грани. Для того чтобы геометрия с разных сторон склеиваемых граней и рёбер была согласована, лучше всего предоставить проведение отождествлений некоторой группе изометрий объемлющего пространства. В 1912 году Гизекинг отождествил грани идеального тетраэдра (то есть с вершинами на абсолюте–поверхности, представляющей бесконечно удаленные точки гиперболического пространства) в гиперболическом пространстве в соответствии с действием некоторой группы изометрий, включающей в себя изометрии, обращающие ориентацию (см. работу Магнуса [17]). Гиперболическое пространство Лобачевского $H^3$ обладает еще одним удивительным свойством, связанным с возможностью разбиения его на правильные пентагональные додекаэдры. Данным вопросом занимались Вебер и Зейферт в работе [18] и установили, что правильный пентагональный додекаэдр является фундаментальной областью кристаллографической группы движений геометрии $H^3$, а «гиперболическое пространство пентагонального додекаэдра» получается факторизацией пространства $H^3$ по подгруппе конечного индекса, а именно по подгруппе индекса 120 в группе симметрий разбиения.

На рис. 1 изображены два различных сценария потери устойчивости полиэдральной формы додекаэдра. Модель, приведенная на рис. 1, *a* эволюционирует в модель рис. 1, *b*, а модель, приведенная на рис. 1, *c* в модель рис. 1, *d*.

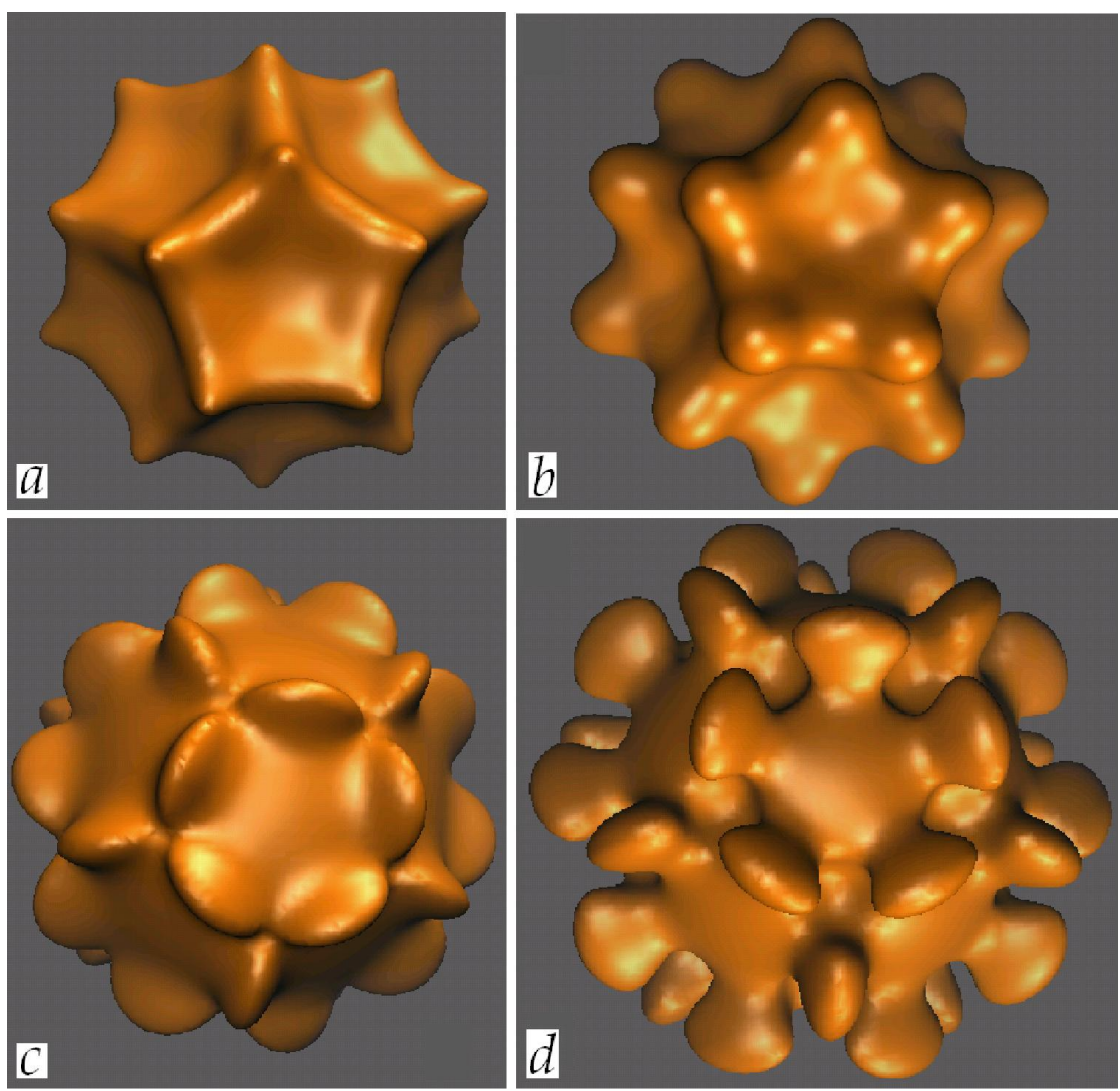

Рис. 1. Два различных сценария потери устойчивости пентагонального додекаэдра: *a* → *b* и *c* → *d*.

Модель приведенная на рис 1, *a* и *b* согласуется с расписанными выше концепциями формирования и роста дендритов, т.е. 20 вершин из-за лучшего питания растут быстрее остальных участков додекаэдра, а в центрах каждого ребра и каждой его грани формируется соответствующий гладкий прогиб.

Из рис. 1, *c* и *d* следует, что возможен также иной сценарий, в котором 30 ребер додекаэдра опережают в росте все остальные его участки. Причем на месте бывших 20 вершин появились характерные прогибы-впадины. Данный сценарий реализуется, когда соответственно ориентированные возмущения появились непосредственно в центрах ребер додекаэдра. Такая эволюция формы возможна тогда, когда подвод вещества к вершинам для их последующего роста существенно затруднен из-за их сложного стехиометрического многокомпонентного состава, а также необходимостью присутствия атомов, у которых малая диффузионная подвижность.

Участки (вершины, грани и ребра) модели дендрита изображенного на рис 1, *a* и *b* – параболоидного типа. Следует отметить, что кривизна поверхности тел изображенных на рис. 1 переменная и принимает значение из интервала [–1; 1]. На данных поверхностях отсутствуют ручки, а также ленты Мёбиуса, следовательно, их род равен нолю, т.е. они гомеоморфные двумерной сфере.

Далее рассмотрены два сценария формирования равновесной формы – усеченный икосаэдр (причем, возможны также и совершенно иные стадии роста). Данные сценарии имеют отличия, связанные с различием в отношениях средней высоты $h$ к средней линейной протяженности $l$ возмущений закономерно возникших на поверхности первоначально-заданного шарообразного кристалла. Отметим, что возмущения, о которых идет речь из-за анизотропных свойств квазирешеток при своем появлении ориентируются определенным образом.

Из рис. 2, *a* следует, что средняя высота $h$ и линейная протяженность $l$ появившихся возмущений в отношении будут $h/l < 1$. Анализируя модели изображенные рис. 2, *b* и *c* можно заключить, что такие возмущения на начальных стадиях развиваются преимущественно тангенциально к поверхности тела.

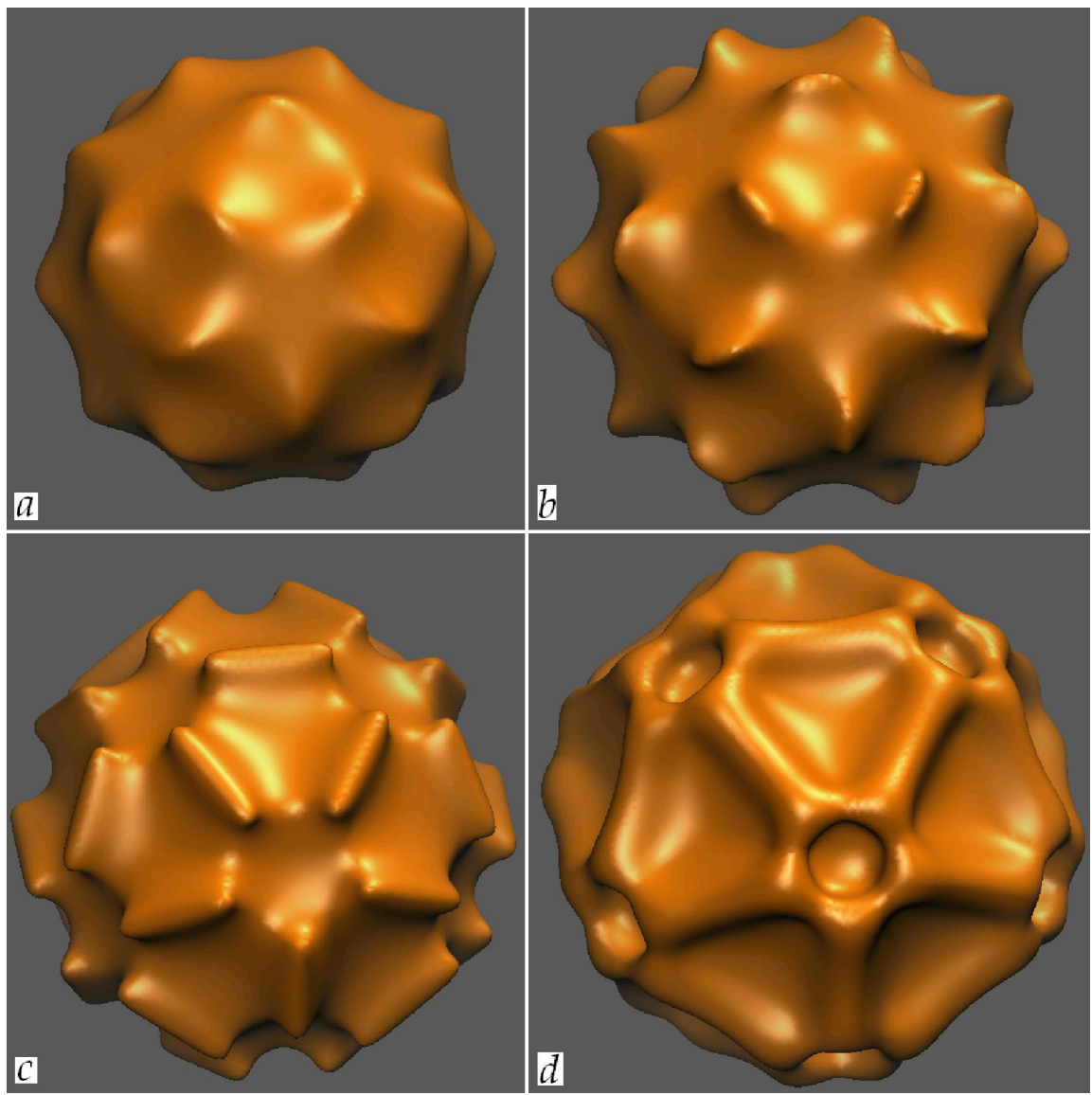

Рис. 2. Последовательные стадии эволюции формы шарообразного кристалла, формирование равновесной формы – усеченный икосаэдр (сценарий 1).

На рис. 2, *c* приведена одна из последовательных стадий роста, когда продольный, тангенциальный рост возмущений (рост вдоль поверхности) почти прекратился и сменился на рост данных возмущений в двух других соответствующих ортогональных направлениях. Из модели, изображенной на рис. 2, *d* следует, что все возмущения соединились между собой, при этом можно сделать ложное заключение, что в процессе роста кристалла прогибы в центрах его граней сформировались под влиянием, например накопления примеси. Другими словами, прогиб на равновесных формах декантированных квазикристаллов может быть следствием довольно сложных процессов их анизотропного роста.

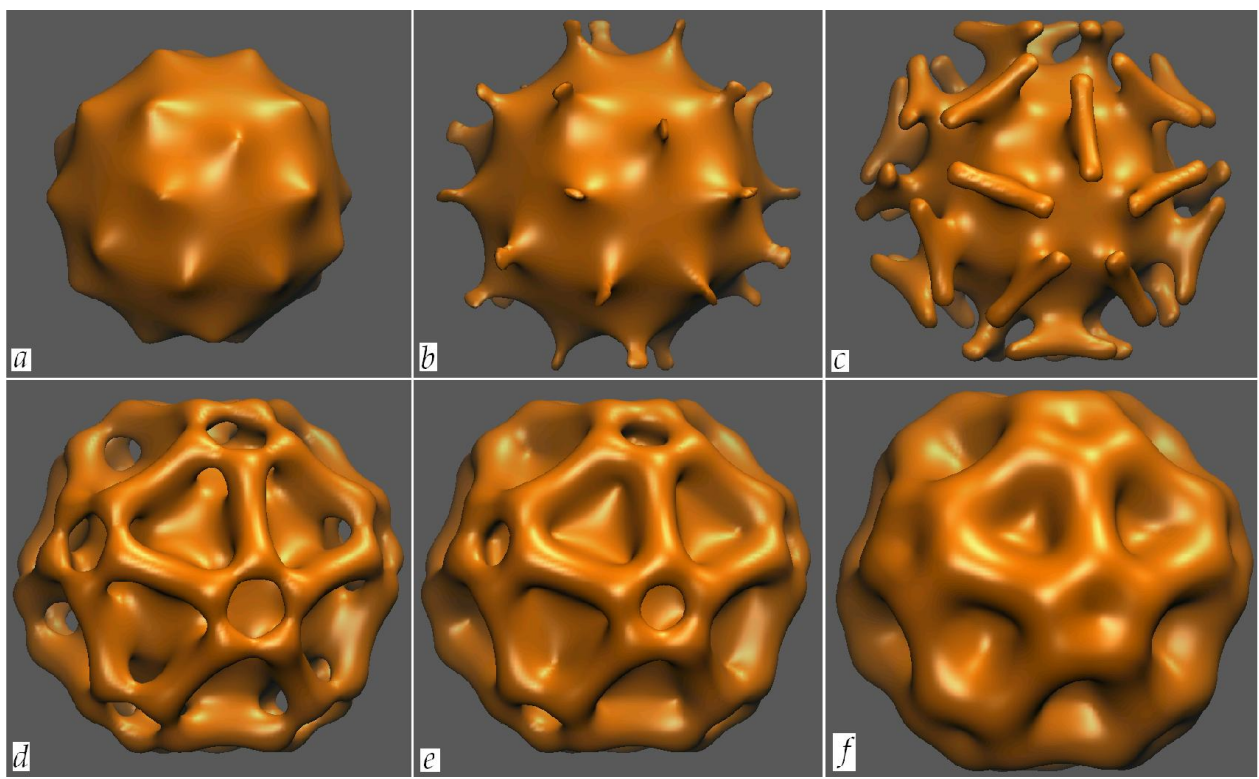

Рис. 3. Последовательные стадии эволюции формы роста шарообразного кристалла, формирование равновесной формы – усеченный икосаэдр (сценарий 2).

На рис. 3 приведена модель последовательных стадий потери устойчивости формы первоначально заданного шарообразного тела, формирование равновесной формы – усеченный икосаэдр, проходя через стадию дендритного квазикристалла. Число симметрично расположенных возмущений появившихся на поверхности тела изображенного рис. 3, *a,* также как и на рис. 2, *a* равно 30. Расположение данных возмущений можно сопоставить с расположением 30 вершин в икосододекаэдре (порядок точечной группы симметрий, которого равен 120).

Из рис. 3, *a* следует, что отношение $h/l > 1$. Из модели, изображенной на рис. 3, *b* видно, что на начальной стадии роста такие возмущения развиваются преимущественно нормально к поверхности тела. Из рис. 3, *c* следует, что на определенной стадии нормальный рост возмущений (имеется ввиду нормальный к поверхности тела) сменяется тангенциальным ростом, образуя при этом *T*-образной формы выступы. Другими словами, на рис. 3, *c* изображена модель дендрита со сформированными в процессе роста симметрично-расположенными боковыми ветвями на каждом его стволе. Заметим, что данные боковые ветви сформировались и начали расти непосредственно из вершин выступов. Далее такие *T*-образные возмущения, разрастаясь как тангенциально, так и утолщаясь, соединяются, образуя поверхность, изображенную на рис. 3, *d*, род которой равен 60.

На рис. 2, *d* и рис. 3, *e* приведена одна и та же равновесная форма кристалла усеченного икосаэдра, у которого в центрах пятиугольных и полуправильных шестиугольных граней сформировались гладкие прогибы-впадины. Отличие в формировании данной равновесной формы заключается в промежуточных стадиях ее образования.

На рис. 4 изображены последовательные стадии развития дендритного квазикристалла род поверхности, которого в процессе роста меняется от 30 до 120 непрерывно путем закономерного формирования ручек. Для лучшей визуализации особенностей строения поверхности дендрита, модели на рис. 4 изображены в разном масштабе. На рис. 4, *a-c* двенадцать симметрически расположенных куполообразных выступов-возмущений совпадают с расположением двенадцати вершин в икосаэдре. Вокруг каждого такого выступа имеется определенное число ручек. На рис. 4, *a* общее число ручек равно 30; на рис. 4, *b* – 60 ручек; на рис. 4, *c* – 120 ручек. Следовательно, род *p* поверхностей изображенных на рис. 4, *a*, *b* и *c* равен 30, 60 и 120 соответственно. Отметим, что данные ручки расположены симметрично на рёбрах «пентагонов» в центрах, которых находятся вышеуказанные возмущения. На рис. 4, *a*, *b* и *c* число таких пентагонов равно 12, 24 и 24 соответственно. Причем, число 24 означает, что, внутри каждого из 12 смежных пентагонов расположен пентагон с меньшей площадью (коэффициент пропорциональности равен числу золотого сечения, τ = 2cos(36°) = 2sin(54°)). Меньшие по площади пентагоны являются обособленными, несмежными.

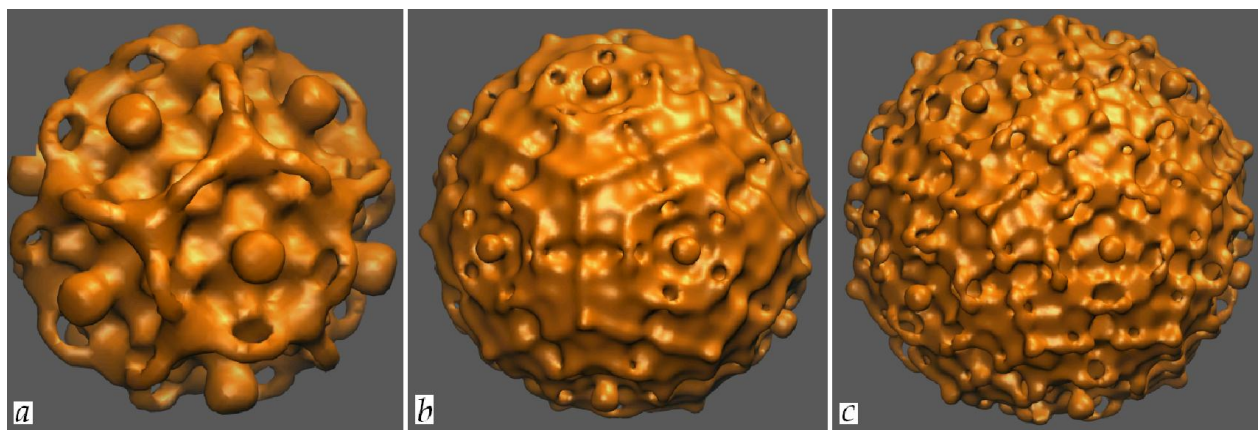

Рис. 4. Модель развития дендритного квазикристалла; на *a*, *b* и *c* род поверхности соответственно равен 30, 60 и 120.

На рис. 4, *a* и *b* положение всех ручек совпадают с центрами рёбер 12 смежных и несмежных пентагонов соответственно. На рис. 4, *c* шестьдесят ручек совпадают с центрами рёбер 12 несмежных пентагонов, остальные 60 ручек расположены по две на расстоянии 1/3 от начала рёбер каждого из 12 смежных пентагонов. Отметим, что если на одной из стадий роста дендритного квазикристалла в определенном месте появилась ручка, то на последующих стадиях на том же месте может сформироваться прогиб.

На рис. 5 приведена модель последовательных стадий роста дендритного квазикристалла с довольно сложным строением поверхности. На рис. 5, *a* симметрическое расположение 12 возмущений куполообразной формы совпадают с 12 вершинами в икосаэдре. Через каждую пару противоположно-расположенных возмущений проходят поворотная ось 5-го порядка и инверсионная ось 10-го порядка. На рис. 5, *a* окружностью белого цвета выделена область, в которой в процессе роста дендрита происходит формирование ручки.

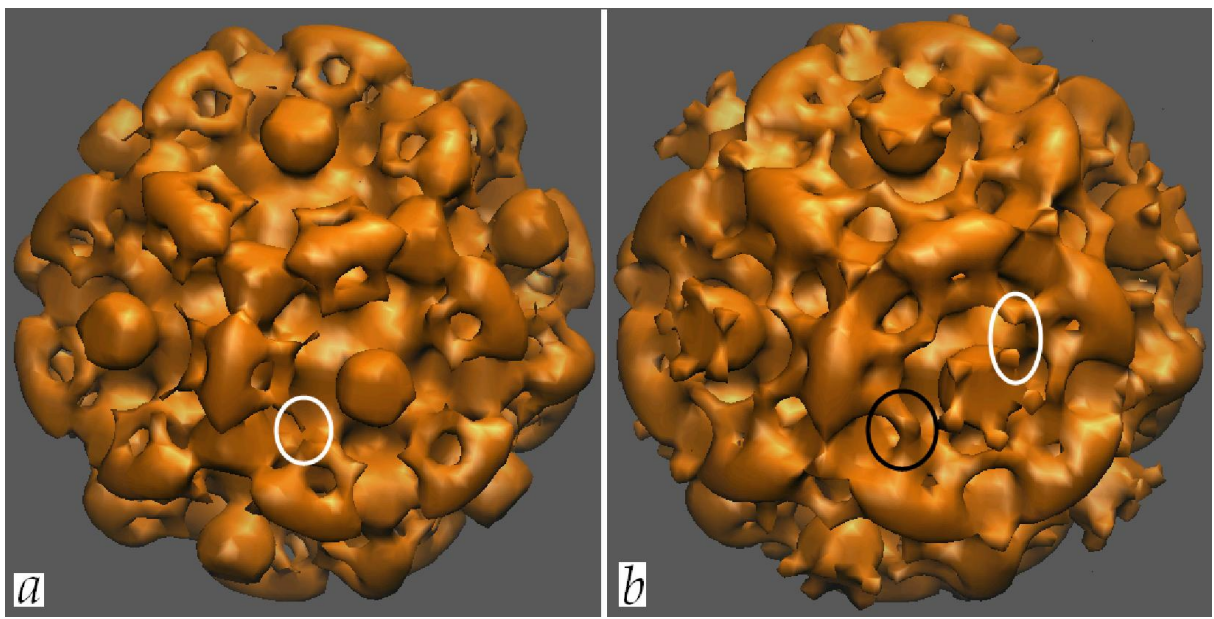

Рис. 5. Последовательные стадии эволюции формы дендритного квазикристалла; на *a* изображена поверхность, род которой $p = 210$, на *b* изображена поверхность, род которой равен 270.

На рис. 5, *b* белой окружностью выделена область, в которой также происходит появление ручки. Данная ручка впоследствии соединит куполообразный выступ с уже имеющейся ручкой на поверхности дендрита (окружность темного цвета), причем каждый такой выступ соединится 5-ю ручками, что добавит 60 ручек к уже имеющимся 270 ручкам. При дальнейшем росте дендрита изображенного на рис. 5, *b* появятся дополнительные 60 ручек и, следовательно, род поверхности его станет 330. Однако не всегда род поверхности в процессе роста дендрита становится численно больше, это связано с тем, что уже имеющиеся ручки на поверхности дендрита, в процессе его роста «зарастают» и визуально исчезают.

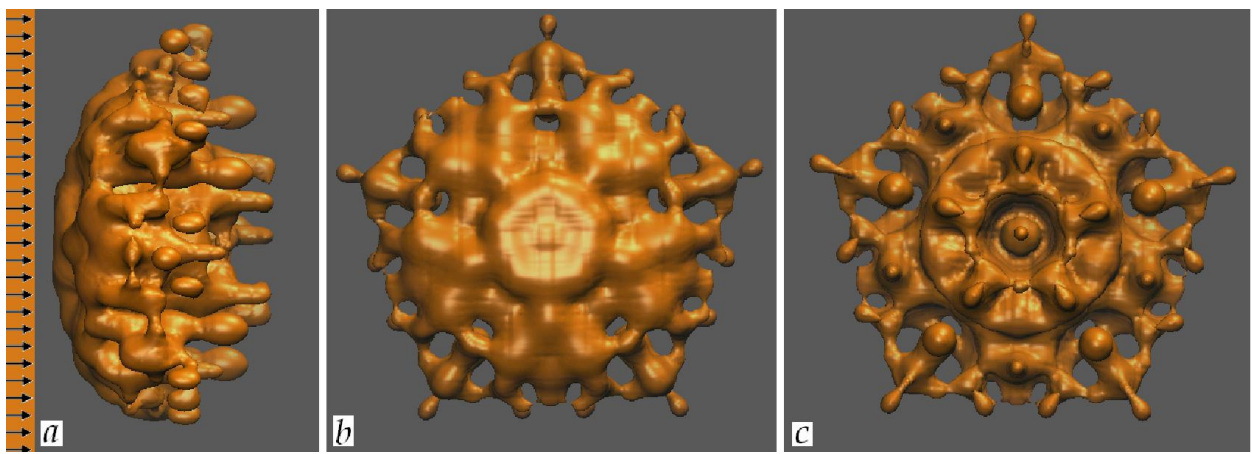

Рис. 6. Модель дендритного квазикристалла (в трех ракурсах) растущего в условиях набегающего потока жидкости; для *a* направление потока указано стрелками; на *b* показана поверхность дендрита, омываемая потоком жидкости; на *c* изображена поверхность дендрита не омываемая потоком жидкости.

Если, например, погрузить в ненасыщенный в среднем раствор кристалл произвольной формы, то одни участки его поверхности будут растворяться, а другие – расти за счет пересыщения, обусловленного поверхностной энергией. Изменение формы кристалла прекратится лишь после установления равновесия над всей его поверхностью. Форма поверхности, обеспечивающая выполнение этого условия, также называется равновесной.

Если же развитие дендрита происходит в условиях набегающего потока расплава или раствора (направление потока совпадает с поворотной осью кристалла), то в зависимости от характеристик потока и условий среды (в которой находится данный дендрит) может, очевидно, происходить с ним следующее: равновесие со средой, растворение или же рост. Из модели, изображенной на рис. 6 следует, что дендрит ускоренно растет вдоль направления потока и сравнительно медленно прорастает в противоположном направлении. Из рис. 6 следует, что из-за наличия потока жидкости ручки на поверхности дендрита сформировались несимметрично и неодновременно. Однако, несмотря на несимметричность данного дендрита, у него сохранились общие признаки наличия поворотной оси 5-го порядка. Таким образом, рост дендритного квазикристалла в условиях набегающего потока жидкости может происходить с сохранением механизма формирования ручек.

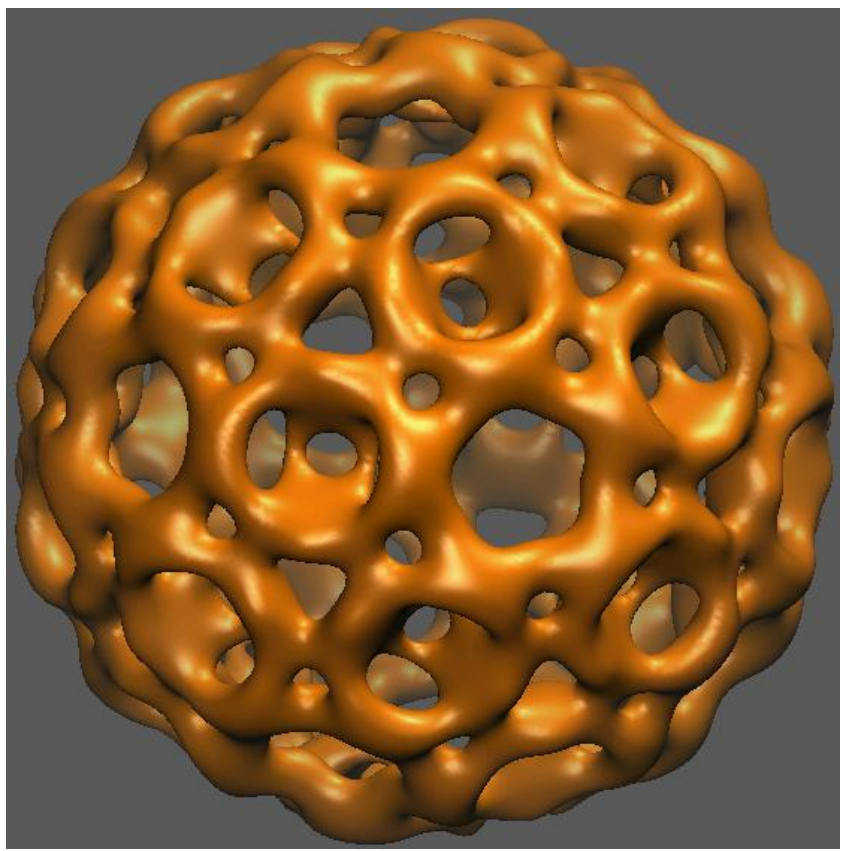

Рис. 7. Модель многосвязной поверхности
обладающей икосаэдральной группой симметрии.

В заключении вспомним одно топологическое свойство орбит в **k**-пространстве. Если поверхность Ферми выходит за границы одной ячейки зонной структуры на ее противоположных гранях или вершинах, то в расширенной зонной схеме поверхность Ферми будет иметь вид многосвязной поверхности, непрерывным образом распределяющейся по всему **k**-пространству. Отдельные результаты данной работы наводят на мысль о некоторой корреляции поверхностей дендритов с возможными поверхностями Ферми в квазикристаллах с икосаэдральной симметрией. При этом можно усмотреть также внешнюю схожесть с некоторыми вирусами, которые обладают икосаэдральной симметрией.

В качестве разминки, читателям предлагается посчитать род поверхности приведенной на рис. 7.

## 3. ЗАКЛЮЧЕНИЕ

Данное исследование посвящено геометрическому моделированию кристаллов. Преимущество такого моделирования в сравнении с известными моделями фазового поля или теоретическим моделированием заключается в том, что отсутствует необходимость решения систем нелинейных дифференциальных уравнений, а также осуществлять подгонку коэффициентов входящих в соответствующие уравнения.

Можно заключить, что сложный анизотропный рост квазикристаллов происходит по характерным закономерностям, которые в свою очередь зависят как от геометрических в частности топологических свойств, так и от состава расплава. Показано, что формирование определенной равновесной формы кристалла может происходить по различным сценариям. Интересной особенностью роста дендритов квазикристаллов, является образование и эволюция ручек на их поверхности.

Неоднородность в атомном строении квазикристаллов свидетельствует о том, что на разных стадиях роста необходима различная по составу доставка питательного вещества к его фронту. Другими словами, атомарный состав меняется закономерно радиально от центра кристалла к его периферии. Вблизи центра симметричного квазикристалла его атомарная плотность численно превышает периферийную (классические кристаллы – однородны, а квазикристаллы – неоднородны). Следовательно, эффект нестабильности некоторых квазикристаллов полученных экспериментально может быть объяснен несогласованностью в их атомном строении. Такая несогласованность влечет за собой накопление напряжений в квазирешетке, что вызывает впоследствии самопроизвольное, неконтролируемое разрушение кристалла.

Впервые смоделирован рост квазикристалла с поворотной симметрией 5-го порядка в условиях набегающего потока расплава. Установлено, что шераховатость поверхности кристалла зависит от его ориентации относительно направления потока жидкости.

## 4. СПИСОК ЛИТЕРАТУРЫ